\documentclass[epsf,graphics,a4,12pt]{article}
\usepackage{epsfig}
\begin{document}
\baselineskip=18pt
\newcommand{\be}{\begin{equation}}
\newcommand{\en}{\end{equation}}
\begin{center}
{\large  Aspects of Higher Order Gravity and Holography}
\end{center}
\vspace{1ex}
\centerline{\large
Elcio Abdalla and L. Alejandro Correa-Borbonet}
\begin{center}
{Instituto de F\'{\i}sica, Universidade de S\~{a}o Paulo,\\
C.P.66.318, CEP 05315-970, S\~{a}o Paulo, Brazil}
\end{center}
\vspace{6ex}
\begin{abstract}
Some thermodynamical properties of Lovelock gravity are discussed in
several space-time dimensions, the holographic principle being one of 
the ingredients of the discussion. As it turns out, the area law and the
brickwall method, though correct for the Einstein-Hilbert theory, may fail
to work in general.  
\end{abstract}
%\vspace{6ex} \hspace*{0mm} PACS number(s): 00000000.
\vfill

%%%%%%%%%%%%%%%%%%%%%%%%%%%%%%%%%%%%%%%%%%%%%%%%%%%%%%%%%%%%%%%%%%%%%%%%%
\section{Introduction.}
%%%%%%%%%%%%%%%%%%%%%%%%%%%%%%%%%%%%%%%%%%%%%%%%%%%%%%%%%%%%%%%%%%%%%%%%%
Since the work of Bekenstein and Hawking \cite{beke,hawking}  our knowlegde
about black hole physics has improved quite considerably. Moreover, black 
hole physics is also the main gate towards understanding of gravity in 
extreme conditions, and as a consequence, of quantum gravity. This led 
t'Hooft and Susskind \cite{thooft,susskind} to generalize the area law
relating entropy and the area of a black hole to any gravitational system
by means of the introduction of the  holographic principle, which
in the last few years turned into a powerful means to the understanding of
possible ways towards the quantization of gravity.

Under such a motivation, the holographic principle was put forward,
suggesting that microscopic degrees of freedom that build up the
gravitational dynamics do not reside in the bulk space-time but on its 
boundary \cite{thooft,susskind}. This principle is a large conceptual
change in our thinking about gravity. Maldacena's conjecture on AdS/CFT 
correspondence \cite{malda} is the first example realizing such a 
principle. Subsequently, Witten \cite{malda} convincingly argued
that the  entropy, energy and temperature of CFT at high temperatures can be
identified with the entropy, mass and Hawking temperature of the AdS 
black hole \cite{hawk-page}, which further supports the holographic 
principle. In cosmological settings, testing the holographic principle is 
somewhat subtle. Fischler and Susskind (FS) \cite{fischsussk}
have shown that for flat and open
Friedmann-Lemaitre-Robertson-Walker(FLRW) universes the area of the
particle horizon should bound the entropy on the backward-looking light
cone. However violation of FS bound was found for closed FLRW
universes. Various different modifications of the FS version
of the holographic principle have been raised subsequently \cite{various}. In
addition to the study of holography in homogeneous cosmologies, attempts
to generalize the holographic principle to a generic realistic
inhomogeneous cosmological setting were carried out in \cite{inho}. 

It is now natural to ask which premises should be forcefully fullfilled
in order to acomodate the holographic principle. In particular, what kind
of dynamics requires holography as an outcome. This could provide a
mechanism for selecting the correct gravity dynamics leading towards
formulating quantum gravity.

The study of the thermodynamic properties of black holes has been 
extended to higher-derivative gravity theories\cite{myers}, known as Lovelock gravity
\cite{lovelock}. Lovelock gravity is exceptional in the sense that
although containing higher powers of the curvature in the Lagrange
density, the resulting equations of motion contain no more 
that second derivatives of the metric. It is also a covariant and ghost free 
theory as happens in the case of Einstein's General Relativity. 

An important result that was found in the thermodynamic context is that
the area  law is a peculiarity of the Einstein-Hilbert theory \cite{scan}. 
This fact motivate us to perform a deeper study of the thermodynamics of 
the black hole solutions of such exotic theories. In \cite{scan} 
gravitation theories are considered with the dimension $d$ and the degree $k$
of the curvature in the respective Lagrangian as parameters. We shall
first briefly review such a formulation and later consider holography in
this context. Further discussions concerning higher derivative gravity can
be found in \cite{serguei}.

%%%%%%%%%%%%%%%%%%%%%%%%%%%%%%%%%%%%%%%%%%%%%%%%%%%%%%%%%%%%%%%%%%%%%%%%%
\section{Higher Dimensional Gravity}
%%%%%%%%%%%%%%%%%%%%%%%%%%%%%%%%%%%%%%%%%%%%%%%%%%%%%%%%%%%%%%%%%%%%%%%%%
The Lanczos-Lovelock action is a polynomial of degre $[d/2]$ in the curvature,
which can be expressed in the language of forms as \cite{scan}
\be
I_{G}=\kappa \int \sum_{m=0}^{[d/2]}\alpha _{m}L^{(m)},  \label{eq:accion}
\en
where $\alpha _{m}$ are arbitrary constants, and $L^{(m)}$ is given by
\be
L^{(m)}=\epsilon _{a_{1}\cdots a_{d}}R^{a_{1}a_{2}}\!\cdot \!\cdot \!\cdot
\!R^{a_{2m-1}a_{2m}}e^{a_{2m+1}}\!\cdot \!\cdot \!\cdot \!e^{a_{d}}.
\label{lp}
\en
$R^{ab}$ are the Riemann curvature two-forms given by
\be
R^{ab}=d\omega^{ab}+\omega^{a}_{c}w^{cb}\quad .
\en
Here $w_{ab}$ are the spin connection one-forms and $e^{a}$ the vielbein. 
A wedge product between forms is understood throughout.

The corresponding field equations can be obtained varying with respect to
$e^{a}$ and $w^{ab}$. In \cite{scan} the expression for the coefficients
$\alpha_{m}$ was found requiring the existence of a unique cosmological
constant. In such a case these theories are described by the action 
\be
I_{k}=\kappa \int \sum_{p=0}^{k}c_{p}^{k}L^{(p)}\;,  \label{eq:Itk}
\en
which corresponds to (\ref{eq:accion})  with the choice
\be
\alpha _{p}:=c_{p}^{k}=\left\{
\begin{array}{ll}
\frac{l^{2(p-k)}}{(d-2p)}\left(
\begin{array}{c}
k \\
p
\end{array}
\right)  & ,%\text{ }
p\leq k \\
0 & ,%\text{ }
p>k
\end{array}
\right.   \label{Coefs}
\en
for the parameters, where  $1 \leq k \leq [(d-1)/2]$.
For a given dimension $d$, the coefficients $c^{k}_{m}$ give rise to a
family of inequivalent theories, labeled by $k$ which represent the
highest power of curvature in the Lagrangian. This set of theories
possesses only two fundamental constants,  $\kappa$ and $l$, related
respectively to the gravitational constant $G_{k}$ and the cosmological
constant $\Lambda$ through 
\be
\kappa =\frac{1}{2(d-2)\Omega_{d-2}G_{k}}\quad ,
\en
\be
\Lambda=-\frac{(d-1)(d-2)}{2l^{2}}\quad .
\en
Since  we are interested in the black hole solutions that are
asymptotically flat we consider the vanishing cosmological constant limit
case. When $l\rightarrow \infty$ the only non-vanishing terms in
Eq(\ref{eq:Itk}) is the 
kth one; therefore the action is obtained from Eq(\ref{eq:accion}) with the 
choice of coefficients 
\be
\alpha _{p}:=\tilde{c}_{p}^{k}=\frac{1}{(d-2k)}\delta _{p}^{k}\; ,
\label{Coefs0}
\en
in which case the  action reads
\be
\tilde{I}_{k}\!=\!\frac{\kappa }{(d-2k)}\!\!\int \!\!\epsilon _{a_{1}\cdots
a_{d}}R^{a_{1}a_{2}}\!\cdot \!\cdot \!\cdot
\!R^{a_{2k-1}a_{2k}}e^{a_{2k+1}}\!\cdot \!\cdot \!\cdot \!e^{a_{d}}\quad .
\label{ActionSL}
\en
Note that for $k=1$ the Einstein action without cosmological Constant is
recovered, while for $k=2$ we obtained the Gauss-Bonnet action,
\be
I_{2}=\frac{(d-2)!\kappa}{(d-4)}\int d^{d}x \sqrt{-g}(-R_{\mu \nu \alpha 
\beta}R^{\mu \nu \alpha \beta}+4R_{\mu \nu}R^{\mu
\nu}-R^{2})\quad . \label{eq:lagran} 
\en
%These theories are far from exotic. Indeed, they are described by the most
%general Lagrangians construted with the same principles as General
%Relativity, that is, general covariance and second order field equations
%for the metric. 
The existence of physical black hole solutions is used as a
criterion to assess the validity of those theories. In the vanishing
cosmological  
constant limit the black hole solution is \cite{scan}  
\be
ds^{2}=-(1-(\frac{r_{h}}{r})^{\gamma})dt^{2}+\frac{dr^{2}}
{1-(\frac{r_{h}}{r})^{\gamma}}+r^{2}d\Omega^{2}_{d-2}\quad , \label{eq:bhl}
\en
where $r_{h}=(2G_{k}M)^{1/(d-2k-1)}$ is the radius of the event horizon and 
\be
\gamma=\frac{d-2k-1}{k}\quad .
\en
The thermodynamic properties of the black holes in higher order
gravity have been studied in various works\cite{myers}. In the case of the black
hole solution (\ref{eq:bhl}) the Hawking  temperature is given by  
\be
T=\frac{\gamma}{4\pi r_{h}}\quad .\label{eq:temp}
\en
Furthermore, using the partition function, obtained from the Euclidean
path integral, the entropy can be calculated leading to the result 
\be
S_{k}=\frac{2\pi k}{G_{k}}\frac{r^{d-2k}_{h}}{d-2k}.\label{eq:entropy}
\en
that is an increasing function of $r_{h}$ which is consistent with the second 
law of thermodynamics.
%%%%%%%%%%%%%%%%%%%%%%%%%%%%%%%%%%%%%%%%%%%%%%%%%%%%%%%%%%%%%%%%%%%%%%%%%%%%
\section{Bounds in Higher order gravity.}
%%%%%%%%%%%%%%%%%%%%%%%%%%%%%%%%%%%%%%%%%%%%%%%%%%%%%%%%%%%%%%%%%%%%%%%%%%%%
Some time ago Bekenstein \cite{entropybound} proposed that exits a
universal upper bound to the entropy-to-energy ratio of any system of
total energy $E$ and effective proper radius $R$ given by the inequality
\be
S/E \leq 2\pi R.
\en
This bound has been checked in many physical situations, either for
systems with maximal  
gravitational effects (i.e strong gravity, such as black holes) or systems
with negligible self-gravity\cite{varios}. 

In this section  we want to consider how this bound behaves
when we consider with the Lovelock 
gravity. First we will obtain the bound for the black hole solutions
(\ref{eq:bhl}).  
Using the entropy relation (\ref{eq:entropy}) and the horizon radius
expression we  
get the bound in an obvious way 
\be
S/E=\frac{\frac{2\pi k}{G_{k}}\frac{r^{d-2k}_{h}}{d-2k}}
{\frac{r^{d-2k-1}_{h}}{2G_{k}}}
=\frac{4\pi k r_{h}}{d-2k}=\frac{2k}{d-2k}(S/E)_{Bek}
\en
We thus obtain that the bound for $S/E$ is $2k/(d-2k)$ times the bound
found by Bekenstein  
for the Schwarzschild case($d=4,k=1$). A real upper bound of $S/E$ for 
these black hole solutions is achieved for the maximal value of the
function $2k/(d-2k)$, namely $d-1 $ for $d$ odd and $\frac{d-2}2$ for $d$
even. In the case of weakly self-gravity systems finding  the bound requires
more steps. We consider a neutral body of rest mass $m$, and proper radius
$R$, that is dropped into the Lovelock type black hole. We also demand
that this process satisfies the generalized second law (GSL).

%Bekenstein bound\cite{beken2}.
%\be
%S/E< C_{0}R
%\en
%(Comentario:Therefore the GSL does not imply the HB for this kind of
%black hole.  
%Here we obtain something more restrictive)

Following Carter \cite{carter} and using the constants of motion(we
consider the metric form  
$ds^{2}=g_{tt}dt^{2}+g_{rr}dr^{2}+r^{2}d\Omega^{2}_{d-2}$)
\begin{eqnarray}
E&=&-\pi_{t}=-g_{tt}\;\dot{t}\\
m&=&(-g_{\alpha \beta}P^{\alpha}P^{\beta})^{1/2}
\end{eqnarray}
we get the equation of motion of the body on the background (\ref{eq:bhl}) 
\be
E=m\sqrt{-g_{tt}}
\en
The energy at $r=r_{h}+\epsilon$ is given by
\be
E=m\gamma^{1/2}(\frac{\epsilon}{r_{h}})^{1/2}
\en
In order to find the change in the black hole entropy caused by the
assimilation of the body, one should evaluate $E$ at the point of capture,
a proper distance $R$ outside the horizon 
\be
R=\int^{r_{h}+\epsilon(R)}_{r_{h}}
\frac{dr}{\sqrt{1-(\frac{r_{h}}{r})^{\gamma}}} 
\en
Integrating we get
\be
R=2\sqrt{\frac{r_{h}\epsilon}{\gamma}}
\en
Therefore we can rewrite the energy as
\be
E=\frac{m\gamma R}{2r_{h}} \quad .
\en
The assimilation of the body results in a change $dM=E$ in the black hole
mass. Using the first law of thermodynamics
\be
dM=T\;dS
\en
and the temperature relation (\ref{eq:temp}) we get that the black hole
entropy increases as 
\be
(dS)_{bh}=2\pi mR
\en
However, we know from GSL, that the relation $(\Delta S)_{T}\equiv
(dS)_{bh}-S_{bo} \geq 0 $ must be  
satisfied. This implies an upper limit for the entropy of the body 
\be
S_{bo}\leq 2\pi ER \label{eq:ubound} \quad .
\en

Once more it is check that the bound (\ref{eq:ubound}) is universal for
negligible self-gravity systems  
because it depends only of the system parameters not of the black hole
parameters. 
%%%%%%%%%%%%%%%%%%%%%%%%%%%%%%%%%%%%%%%%%%%%%%%%%%%%%%%%%%%%%%%%%%%%%%%%%%%%
\section{Brick Wall Method.}
%%%%%%%%%%%%%%%%%%%%%%%%%%%%%%%%%%%%%%%%%%%%%%%%%%%%%%%%%%%%%%%%%%%%%%%%%%%%
%El metodo de brick wall no diferencia cual potencia tiene r/r_{h} en el 
%factor de horizonte 
Another interesting point is to check the method of brick wall\cite{hooft} for this
kind of black holes. As an example we perform the calculations for black
holes in $d=8$ and $k=2$. Therefore, we have
\be
ds^{2}=-hdt^{2}+h^{-1}dr^{2}+r^{2}d\Omega^{2}_{4} \quad , 
\en
where the function $h(r)$ function which describes the event horizon,
is given by, 
\be
h=1-(\frac{r_{h}}{r})^{3/2}\quad .
\en
In this background, we consider a minimally coupled scalar field which
satisfies the Klein-Gordon equation\cite{birell}
\be
\frac{1}{\sqrt{-g}}\partial_{\mu}(\sqrt{-g}g^{\mu \nu}\partial_{\nu}\Phi)-
m^{2}\Phi =0\quad . \label{eq:gordon}
\en
The 't Hooft method consists in introducing a brick wall cut-off near the
event horizon, such that the boundary condition
\be
\Phi=0 \;\;\;\;\; for \;\;\;\; r\leq r_{h}+\epsilon \quad 
\en
is satisfied. In order to eliminate infrared divergencies, another cut-off
is introduced at a large distance from the horizon, $ L \gg r_{h} $, where
we have, 
\be
\Phi=0 \;\;\;\;\; for \;\;\;\; r\geq L \quad .
\en 
In the spherically symmetric space, the scalar field can be decomposed as 
\be
\Phi(t,r,\theta)=e^{-iEt}R(r)Y(\theta) \quad ,
\en
where $\theta$ represents all the angular variables.
Substituting this expression back into (\ref{eq:gordon}) and using the
eigenvalue equation for the generalized spherical function $Y(\theta)$,
\be
\triangle \; Y(\theta) =-l(l+5)\; Y(\theta) \quad , 
\en 
we obtain, after some manipulations, the radial equation
\be
h^{-1}E^{2}R(r)+\frac{1}{r^{6}}\partial_{r}[r^{6}h\partial_{r}R(r)]-
\frac{l(l+5)}{r^{2}}R(r)-m^{2}R(r)=0 \label{eq:radi}\quad .
\en
Using the WKB approximation, we substitute $R(r)=\rho(r)e^{iS(r)} $, the
function $\rho(r)$ being   
a slowly varying amplitude and $S(r)$ is a rapidly varying phase. To
leading order, only the  
contribution from the first derivatives of $S$ are important. Then from
eq (\ref{eq:radi}) we  
get for the radial wave number $K\equiv \partial_{r}S$, the expression
\be
K=\left(1-(\frac{r_{h}}{r})^{3/2}\right)^{-1}\sqrt{E^{2}-
\left(1-(\frac{r_{h}}{r})^{3/2}\right)
\left(\frac{l(l+5)}{r^{2}}+m^{2}\right)} \quad .
\en
In such a case, the number of radial modes $n_{r}$ is given by
\be
\pi n_{r}=\int^{L}_{r_{h}+\epsilon}dr K(r,l,E) \quad . \label{eq:modes}
\en
In order to find the entropy of the system we calculate the free energy
of a thermal bath of scalar particles with an inverse temperature $\beta$,
that is 
\be
e^{-\beta F}=\sum e^{-\beta E_{N_{\tau}}} \quad , 
\en
where $E_{N_{\tau}}$ is the total energy corresponding to the quatum state
$\tau$. Since the sum also includes the degeneracies of the quantum, we have
\be
e^{-\beta F}=\prod_{n_{\tau}} \frac{1}{1-exp(-\beta E)} \quad ,
\en 
where $(n_{\tau})$ represents the set of quantum numbers associated to this
problem. The product  
$\prod $ take into account the contribution from all the modes. The factor
$(1-e^{-\beta E})^{-1}$ is due  
to the fact that we are dealing with bosons where the ocupation number can
take on the value of  
all positive integers as well as zero, so that
\be
\sum^{\infty}_{n=0} e^{-\beta nE}=\frac{1}{1-exp(-\beta E)} \quad .
\en 
From the previous equation we can write the free energy as
\begin{eqnarray}
F & = & \frac{1}{\beta} \sum log(1-e^{-\beta E}) \nonumber \\
  & = & \frac{1}{\beta} \int dl D_{l}\int dn_{r} log(1-e^{-\beta E})
\end{eqnarray}
where 
\be
D_{l}=\frac{(2l+5)(l+4)!}{5!l!}=\frac{(2l+5)(l+1)(l+2)(l+3)(l+4)}{5!}
\en
is the degeneracy of the spherical modes\cite{bateman}.

Integrating by parts and using (\ref{eq:modes}) we get
\begin{eqnarray}
F & = & -\int dl \;D_{l} \int dE \frac{1}{exp(\beta E)-1}n_{r} \nonumber \\
  & = & -\frac{1}{\pi} \int dl \;D_{l} \int dE \frac{1}{exp(\beta E)-1} 
\int^{L}_{r_{h}+\epsilon} dr \nonumber \\
  &  & \times \left(1-(\frac{r_{h}}{r})^{3/2} \right)^{-1} 
  \sqrt{E^{2}-\left(1-(\frac{r_{h}}{r})^{3/2} \right)\left(\frac{l(l+5)}{r^{2}}
  +m^{2}\right)} 
\end{eqnarray}
The $l$ integration can be perfomed explicitly and it is taken only over those 
values for which the square roots exits,
\begin{eqnarray}
\lefteqn{\int dl \; D_{l} \;\sqrt{E^{2}-\left(1-(\frac{r_{h}}{r})^{3/2}
  \right)\left(\frac{l(l+5)}{r^{2}} 
  +m^{2}\right)}= }\nonumber\\
 & & \frac{16r^{6}(E^{2}-hm^{2})^{7/2}}{5!105h^{3}}+\frac{8r^{4}(E^{2}-
hm^{2})^{5/2}}{5!3h^{2}}+\frac{16r^{2}(E^{2}-hm^{2})^{3/2}}{5!h} 
\end{eqnarray}
We are interested in the leading contribution to the free energy near the
horizon. Then we just take the first term from the previous equation,
that is, 
\be
F=- \frac{16}{5!105\pi}\int dE \frac{1}{exp(\beta E)-1}\int^{L}_{r_{h}
+\epsilon} dr r^{6} h^{-4}
\left[E^{2}-hm^{2}\right]^{7/2}\label{eq:free} \quad .
\en  
Introducing the change of variable $y=(\frac{r}{r_{h}})^{3/2}$ and
substituting it back into (\ref{eq:free}) we find
\be
F=- \frac{32 r^{7}_{h}}{5!315\pi}\int dE \frac{1}{exp(\beta E)-1}
\int^{\bar{L}^{3/2}}_{(1+\bar{\epsilon})^{3/2}} 
dy y^{11/3}(1-\frac{1}{y})^{-4}\left[E^{2}-(1-\frac{1}{y})m^{2}\right]^{7/2}
\en
where $\bar{\epsilon}=\frac{\epsilon}{r_{h}}$, $\bar{L}=\frac{L}{r_{h}}$.

Near the horizon, that is for $y$ near $1$, we find the expression\cite{mavroma}
\be
F =-\frac{32r^{7}_{h}}{5!315\pi}\int^{\infty}_{0} dE
\frac{E^{7}}{exp(\beta E)-1} 
\int^{\bar{L}^{3/2}}_{(1+\bar{\epsilon})^{3/2}} dy (y-1)^{-4}\quad .
\en 
We next use the formula
\be
\int^{\infty}_{0} dE \frac{E^{7}}{exp(\beta E)-1}=\frac{7!\zeta(8)}{\beta^{8}}
\en
and integrate over $y$. The expression for $F$ reduces to
\be
F=-\frac{2^{9}\zeta(8)}{45\pi
  3^{3}}\frac{r^{10}_{h}}{\epsilon^{3}\beta^{8}} \quad , 
\en
allowing us to compute the entropy from
\be
S=\beta^{2}\frac{\partial F}{\partial \beta}=\frac{2^{12}\zeta(8)}
{45\pi 3^{3}}\frac{r^{10}_{h}}{\epsilon^{3}\beta^{7}} \quad .
\en
The inverse of the Hawking temperature is
\be
\beta =\frac{8\pi}{3} r_{h}
\en
and we can subsequently find the entropy, that is,
\be
S=\frac{3^{4}\zeta(8)}{45\pi^{8}2^{9}}\frac{r^{3}_{h}}
{\epsilon^{3}} \quad .
\en
This expression can be transformed making use of the invariant distance
\be
\int ds=\int^{r_{h}+\epsilon}_{r_{h}}dr \frac{1}{\sqrt{1-(r_{h}/r)^{3/2}}}=
\sqrt{\frac{8r_{h}\epsilon}{3}}\quad , 
\en
in terms of which we can rewrite the entropy as a function of 
invariants only, 
\be
S=\frac{\zeta(8)}{15\pi^{8}}\frac{r^{6}_{h}}{(r_{h}\epsilon)^{3}}=
\frac{A}{D^{(2)}_{(8)}(\int ds)^{6}}\quad , \label{eq:entro2}
\en
where $A=\frac{16}{15}\pi^{3} r^{6}_{h}$ is the horizon area and
$D^{(2)}_{(8)}= \frac{2^{4}\pi^{11}}{\zeta(8)}$.

Therefore the entropy of the scalar field is proportional to the area and
diverges cubically  with the cutoff $\epsilon$. 

In reference \cite{sussuglu} it was shown that the question of finiteness
of the entropy can be solved by the renormalization of the Newton's
gravitational constant. Here that is not possible because the bare entropy
(\ref{eq:entropy}) does not have the same power of the horizon radius as  the
divergent term (\ref{eq:entro2}).  

%Free energy(general result)
Repeating the same procedure we can find the general expression for the
free energy, for given values of $d$ and $K$, which are
\be
F^{(d)}_{\epsilon}=-C^{(k)}_{(d)}\frac{r^{\kappa_{d}}_{h}}
{\epsilon^{\frac{d-2}{2}}\beta^{d}}\quad ,\label{eq:enerfree}
\en
where $\kappa_{d}=d+\frac{d-4}{2}$.

From the previous equation the entropy can also be obtained  also in
an easy way, 
\be
S^{(d)}=\frac{dC^{(k)}_{(d)}\gamma^{d/2}2^{d-2}}{(4\pi)^{d-1}}
\frac{r^{d-2}}{(\int  ds)^{d-2}}= 
\frac{A}{D^{(k)}_{(d)}(\int ds)^{d-2}}\quad ,
\en
where $A=\frac{2\pi^{(d-1)/2}}{\Gamma((d-1)/2)} r^{d-2} $ and
$D^{(k)}_{(d)}=\frac{2^{d+1}\pi^{3/2(d-1)}} 
{dC^{(k)}_{(d)}\gamma^{d/2}\Gamma((d-1)/2)}$.

This result implies  that the brick wall method works just for linear gravity. 

\section{Conclusions.}

In this paper we have studied some properties of the black hole solutions in 
higher order gravity. One of the main conclusions from this study is that we 
can not infer the holographic bound from the Generalized Second
Law (GSL). In other words, the area law is not respected despite the fact that 
the second law of thermodinamics is satisfied. Another interesting outcome is 
that the brick wall method works well only  for the Einstein-Hilbert
theory ($k=1$). A possible explanation is that this method, by
construction, computes the modes living in a shell and therefore at the
end of the calculations always reflects this geometrical set-up.
  
%In order to see  the extension of this results to cosmology we will study the 
%the action (\ref{eq:lagran}) also known as Lanczos action \cite{madore} 
%\be
%ds^{2}=-dt^{2}+a^{2}(t)dx^{i}dx^{i}+\phi^{2}(t)dx^{l}dx^{l}
%\en
%the indices $i$,$l$ take the values $1,2,3$ and $4,5,6,7$ respectively. 
%\be
%a(t)=a_{0}(\frac{t}{t_{i}})^{\alpha} \;\;\;\;\;\; 
%\phi(t)=(\frac{t}{t_{0}})^{\beta}
%\en
%Two solutions
%First
%The standard $k=0$ solution
%Second
%\be
%\alpha=0.53 \;\;\;\;\;\;\;\;\;  \beta=1-3\alpha=0.58
%\en
%\be
%S/A=\frac{(R_{H,a})^{3}(R_{H,b})^{4}}{[(aR_{H,a})^{3}(bR_{H,b})^{4}]^{6/7}}
%\en
%\be
%S/A=t^{-3\beta}
%\en
ACKNOWLEDGMENT:
This work was partially supported
by Funda\c{c}\~ao de Amparo \`a Pesquisa do Estado de
S\~ao Paulo (FAPESP), Conselho Nacional de Desenvolvimento  
Cient\'{\i}fico e Tecnol\'{o}gico (CNPq) and CAPES (Ministry of Education,
Brazil). L.A.C.B thanks the partial support of the High Energy Group of 
the Abdus Salam ICTP where part of this work was perfomed.

{\bf APPENDIX.}

Here we show the value of the constants $C^{(k)}_{(d)}$ in the free energy
expression  
(\ref{eq:enerfree}) for the different values of the dimension $d$ and the
degree  in curvature $k$.
\begin{eqnarray*}
C^{(2)}_{(6)}&=&\frac{64\zeta(6)}{3\pi}\\
C^{(2)}_{(7)}&=&\frac{3\zeta(7)}{4}\\
C^{(3)}_{(8)}&=&\frac{24 3^{3}\zeta(8)}{5\pi}\\
C^{(2)}_{(9)}=\frac{5\zeta(9)}{2^{9/2}16} \;\;\;\;&&\;\;\;\; 
C^{(3)}_{(9)}=\frac{5\zeta(9)}{16}\\
C^{(2)}_{(10)}=\frac{2^{11}\zeta(10)}{7\pi 5^{6}}\;\;\;\;\;\;\;\;
C^{(3)}_{(10)}&=&\frac{8^{2}\zeta(10)}{35\pi}\;\;\;\;\;\;\;\;
C^{(4)}_{(10)}=\frac{2^{16}\zeta(10)}{35\pi}
\end{eqnarray*}
%\newpage
%%%%%%%%%%%%%%%%%%%%%%%%%%%%%%%%%%%%%%%%%%%%%%%%%%%%%%%%%%%%%%%%%%%%%%%%%%%%

\end{document}